\documentclass[aps,preprint]{revtex4-1}

\usepackage{epsf}
\usepackage{epsfig}
\usepackage{amssymb}
\usepackage{bm}
\usepackage{hyperref}
\usepackage{color}
\newcommand{\be}{\begin{equation}}
\newcommand{\ee}{\end{equation}}
\newcommand{\bea}{\begin{eqnarray}}
\newcommand{\eea}{\end{eqnarray}}

\newcommand{\nn}{\nonumber}

\newcommand{\commute}[2]{\left[ #1,#2 \right]}
\newcommand{\anticommute}[2]{\left\lbrace #1,#2 \right\rbrace}

\newcommand{\degree}{^\circ}
\newcommand{\loge}{\log_{10}(E_\mu/\text{GeV})}
\newcommand{\sci}[2]{#1$\times$10$^{\text{#2}}$}
\newcommand{\ck}[1]{\textcolor{black}{#1}}

\begin{document}

\title{Dark matter at DeepCore and IceCube}
\author{V.~Barger$^{1}$, Y.~Gao$^{2}$, D.~Marfatia$^{3,1}$\\[2ex]
\small\it $^{1}$Department of Physics, University of Wisconsin, Madison, WI 53706, U.S.A.\\
\small\it $^{2}$Department of Physics, University of Oregon, Eugene, OR 97403, U.S.A.\\
\small\it $^{3}$Department of Physics and Astronomy, University of Kansas, Lawrence, KS 66045, U.S.A.}
\begin{abstract}
With the augmentation of IceCube by DeepCore, the prospect for detecting dark matter 
annihilation in the Sun is much improved. To complement this experimental development,
we provide a thorough template analysis of the particle physics issues that are necessary to
precisely interpret the data. Our study is about nitty-gritty and is intended as a framework for 
detailed work on a variety of dark matter candidates. To accurately predict the source neutrino spectrum, we account for spin correlations of the final state particles  and the helicity-dependence of their decays, and absorption effects at production. We fully treat the propagation of neutrinos through the Sun, including neutrino oscillations,
energy losses and tau regeneration. We simulate the survival probability of muons produced in the Earth by using the Muon Monte Carlo program, reproduce the published IceCube effective area, and
update the parameters in the differential equation that approximates muon energy losses. To evaluate the zenith-angle dependent atmospheric background event rate, we track the Sun and determine the time it spends at each zenith angle. Throughout, we employ neutralino dark matter as our example.

\end{abstract}

\maketitle

\section{Introduction}

Astrophysical data require about 20\% of the energy density of our universe to be in the form
of unseen matter.  The nature of this {\it dark matter} (DM) is a mystery that is key to resolving several problems in astrophysics and cosmology. 
A possibility is that DM consists of stable or very long-lived Weakly Interacting Massive Particles (WIMPs). However, there is no such particle in the spectrum of the Standard Model (SM) of particle physics and a wide range of new physics models have been proposed to introduce WIMP candidates. The most popular scenarios include supersymmetry (SUSY), extra dimensions, and $n$-plet extended models ($n\geq 1$). 

SUSY with \ck{$\cal{R}$-parity~\cite{bib:susyR} is} a well motivated possibility that alleviates the 
hierarchy problem, realizes gauge coupling unification and facilitates the seesaw mechanism of neutrino mass generation when augmented with right-handed neutrinos. $\cal{R}$-parity conserving SUSY provides the lightest $\cal{R}$-odd particle as a 
WIMP candidate. The cosmic microwave background data from Wilkinson Microwave Anisotropy Probe (WMAP)~\cite{Jarosik:2010iu} pins down the relic dark matter abundance to be 
$\Omega_{DM}h^2=0.1123\pm0.0035 $, which stringently constrains the SUSY parameter space. \ck{The popular minimal supergravity model~\cite{bib:msugra} (mSUGRA)} is defined by a set of only five parameters. 

Indirect searches look for cosmic ray excesses in the diffuse background or from point sources due to DM annihilation or decay into SM particles. 
Due to their high penetration ability, neutrinos help with detecting DM deep inside gravitational wells that include nearby sources like \ck{the Sun, the Earth and the galactic center~\cite{bib:sun,Press:1985ug,bib:earth}}.
\ck{DM-induced neutrinos from the Sun} can be observed if the signal rate is competitive with the atmospheric neutrino flux, which is created by collisions of cosmic protons and nuclei in the atmosphere and is the dominant background below a TeV.

\begin{table}
\begin{center}
\begin{tabular}{l|cccc|c|cccccc|c}
\hline
Point& $m_0$ & $m_{1/2}$ & $A_0$ & $\tan{\beta}$ &$m_{\chi^0}$ &$\tau^+\tau^-$ & $W^+W^-$&$ZZ$&  $b\bar{b}$&$c\bar{c}$&$t\bar{t}$ & \ck{Ann./yr}\\
\hline
A (Focus Point)& 2154 & 288 &0 &10 &105 & \text{---} & \text{90$\%$} & \text{8.4$\%$} & \text{1.0$\%$} & \text{0.11$\%$} & \text{---} & \sci{5.4}{22} \\
B & 2268 & 488 &0 &50 & 197 & \text{1.3$\%$} & \text{12$\%$} & \text{5.4$\%$} & \text{9.5$\%$} & \text{---} &
   \text{69$\%$} & \sci{4.4}{21} \\
C \ck{($\tilde{\tau}\ $co-ann.)}&54 &241 &0 &10 &93 & \text{16$\%$} & \text{4.4$\%$} & \text{---} & \text{76$\%$} & \text{---} &
   \text{---} & \sci{2.3}{21} \\
D ($A$-funnel)& 483 & 304 &0 &50 & 123 & \text{12$\%$} & \text{---} & \text{---} & \text{88$\%$} & \text{---} &
   \text{---}  & \sci{2.7}{21}\\
\hline
\ck{E ($\tilde{t}\ $co-ann.)}&150 &302 &-1099 &5 & 121 &95\%&0.24\%&---&2.9\%&---&--- &$1.2\times 10^{18}$\\ 
F (Bulk)&80 &170 &-250 &10 & 64&36\%&---&---&63\%&---&---&$7.7\times 10^{21}$\\ 
G ($h$-funnel)&2000 &130 &-2000 &10 & 55 &7.4\%&---&---&83\%&3.5\%&---&$4.4\times 10^{16}$\\
\hline
\end{tabular}
\end{center}
\caption{Sample points in regions of mSUGRA parameter space that are compatible with the dark matter relic abundance; the sign of the $\mu$ parameter is positive and the top quark mass is 172.7 GeV.
The lightest neutralino $\chi^0$ is the WIMP candidate. Points A$-$D are selected to have 
$m_{\chi^0}\sim 10^2$~GeV and large annihilation rates from parameter scans with $\tan\beta=10$ and 50. Points E$-$G are representative relic density compatible points from 
Ref.~\cite{Barger:2009gc}.  Masses are in GeV, and the last column gives the number of annihilations per year in the Sun. Only channels with branching fractions larger than $10^{-3}$ are listed. 
}
\label{tab:susy}
\end{table}


\ck{Among the many experiments searching for high energy neutrinos,
 we focus on the IceCube (IC) detector~\cite{icecube} that is capable of observing neutrinos with energies above $100$~GeV.} The installment of DeepCore~\cite{Wiebusch:2009jf} (DC) significantly lowers IceCube's energy threshold and enhances the ability of detecting neutrinos from light WIMP annihilation.
We carry out a detailed simulation of IC/DC detection of the neutrino signal from neutralino annihilation in the Sun for the sample relic-density-consistent mSUGRA 
points listed in Table~\ref{tab:susy} in standard notation; for a review of the various parameter
regions see Ref.~\cite{azar}.

In Section~\ref{sec:solarDM} we summarize  the physics of DM condensation and annihilation in the Sun. In Section~\ref{sec:spec} we describe our calculation of the spin correlated neutrino source spectrum and its propagation from the Solar core to the Earth. In 
 Section~\ref{sec:detector} we discuss the simulation of neutrino-induced events at the IC/DC detector.
We present our results in Section~\ref{sec:signals} and summarize in Section~\ref{sec:res}. In 5 appendices, we provide several 
details of our calculations.

\section{DM capture and annihilation}
\label{sec:solarDM}

As the Sun sweeps through the dark matter halo, WIMPs collide with solar nuclei and become gravitationally trapped. 
\ck{The capture over a long period of time leads to condensation of low-speed WIMPs in the center of the Sun. The capture rate~\cite{Press:1985ug} $C_C$ receives contributions from spin-independent (SI) and spin-dependent (SD) scattering between WIMPs and nuclei. Then,  $C_C=C_C^{SI}+C_C^{SD}$ with~\cite{Jungman:1995df}
}

\bea
C_C^{SI}&=&4.8\times10^{28}s^{-1} \frac{\rho_{0.3}}{\bar{v}_{270} m_{\chi^0}} \sum_{i} F_i f_i \phi_i  \frac{\sigma^{SI}_i }{m_{N_i}}
S\left({\scriptsize \frac{m_{\chi^0}}{m_{N_i}}}\right)\,,\\ 
C_C^{SD}&=&1.3\times10^{29}s^{-1}\frac{\rho_{0.3}}{\bar{v}_{270} m_{\chi^0}}\sigma^{SD}_{H} S\left(  \frac{m_{\chi^0}}{m_{N_i}}\right)\,,
\label{eq:ann}
\eea
where $i$ sums over the elements with significant abundance in the Sun ranging from hydrogen to iron. $\rho_{0.3}$ is the local DM halo density in units of 0.3 GeV/cm$^{3}$, $\bar{v}_{270}$ is the average DM dispersion velocity in units of 270 km/s, $m_{N_i}$ denotes the mass of the nucleus of the $i^{th}$ element in GeV, and $\sigma_{i}$ is the SD/SI  scattering cross section in pb. $f_i, F_i$ and $S$ are the mass fraction, kinematic suppression and form-factor 
suppression~\cite{Kamionkowski:1991nj} for nucleus $i$, respectively. $\phi_i$ describes the distributions of the $i^{th}$ element. We refer interested readers to Ref.~\cite{Jungman:1995df} for a detailed discussion and the values for these parameters. 
For most mSUGRA points consistent with the measured relic density, $\sigma^{SD}$ is 
greater than $\sigma^{SI}$ by two to three orders of magnitude, but does not necessarily dominate the capture rate. 


As the density builds up in the center of the Sun the annihilation of DM particles occurs more frequently. Eventually equilibrium sets in, $C_{C}=2C_{A}$, where $C_A$ is the annihilation rate. 
\ck{
However, it was pointed out in Ref.~\cite{bib:ellis} that large areas of SUSY parameter space do not saturate this equilibrium condition and $C_A$ can be significantly below $C_C/2$. 
 The DM annihilation rate in the Sun~\cite{Griest:1986yu} and can be parametrized by~\cite{Jungman:1995df}}
\be
\begin{array}{ccl}
C_A&=&\frac{C_C}{2}\text{tan}^2\left(
t/\tau
\right)\,,\\

t/\tau &=& 330\left[
\frac{C_C}{\text{s}^{-1}}\frac{\left<\sigma_A v\right>}{\text{cm}^3 \text{s}^{-1}}\left(\frac{m_{\chi^0}}{\text{10 GeV}}\right)^{0.75}
\right]^{\frac{1}{2}}\,,
\end{array}
\label{eq:cap}
\ee
where $t$ and $\tau$ denote the age of the Sun and the equilibrium time scale, and
$\left<\sigma_A v\right>$ is the annihilation cross section averaged over the velocity distribution in the nonrelativistic limit. We do not assume that equilibrium holds, and calculate the annihilation
rate for each region.

\section{Neutrino source spectra and propagation}
\label{sec:spec}

The source neutrino/antineutrino flux is
\be 
\frac{d\phi_{\nu}}{dE_\nu}=C_{A} \sum_{i}\text{BF}_i \frac{d\phi^i_\nu}{dE_\nu}\,,
\label{eq:source}
\ee
where $i$ denotes each annihilation channel, and BF$_i$ and $\frac{d\phi^i_\nu}{dE}$ are the corresponding branching fractions and normalized (to each annihilation event) neutrino energy spectra, respectively.

The dominant annihilation channels for our sample mSUGRA points are provided in 
Table~\ref{tab:susy}. Note that the evolution of the mSUGRA renormalization group equations (RGEs) to the weak scale can be numerically sensitive to the GUT-scale parameters and lead to significant uncertainty in the annihilation rate. This is especially true for the Focus Point region. We use DarkSusy~\cite{Edsjo:1997hp} to calculate $\left<\sigma_A v\right>$ and 
$\sigma^{SI,SD}_{p,n}$, needed to determine the annihilation rate.

Since neutralinos are Majorana fermions, their  annihilation into light fermion pairs is helicity-suppressed in the nonrelativistic limit. Thus WIMP-induced neutrinos do not have a line spectrum in mSUGRA. The dominant neutrino source is the annihilation into gauge bosons,  2$^{nd}$ and $3^{rd}$ generation fermions, and their subsequent decays. Spin-correlation effects are visible in the neutrino energy spectrum, especially in the case of gauge bosons  for which final states with transverse polarizations dominate in the static limit. 
It is important to carry out a spin-correlated calculation since the transversely polarized $WW$ channel produces a significantly harder neutrino spectrum than the longitudinal $WW$ channel. 
Similarly, a left-handed $\tau$ produces more neutrinos than a right-handed $\tau$.
In our analysis, we retain the spins of particles that directly result from the decay of particles pair-produced in DM annihilation. 
Secondary neutrinos arise from subsequent decays, and we include the spin correlation in helicity-dependent $\tau$ decays.  
At this level the spin information of the primary and major secondary neutrino contributions are taken into account.
The $s$-channel top-pair final state can be significant for large neutralino masses, and the polarization of the on-shell $W$ from top decay affects the leading neutrino distribution. For this particular channel we proceed to the next level and keep the spins of the decay products of the $W$. 
See Appendices~\ref{app:spin} and~\ref{app:tau_decay} for a discussion of our simulation of the various annihilation channels.

The dense solar matter absorbs all the muons and relatively long-lived hadrons. A large fraction of $c,b$ hadrons also scatter before they can decay. Thus muons are considered the end of the cascade and do not contribute to the neutrino flux. The absorption probability of $c, b$ hadrons is determined by the competition between scattering and decay rates, which is discussed in Appendix~\ref{app:absorption}.

The low flux density at the center of the Sun means neutrinos are created incoherently in the flavor basis. Following Ref.~\cite{Strumia:2006db}, we treat the neutrino propagation through the Sun 
with the flavor-density matrix $\bm{\rho}_{ij}$ that denotes the distribution in the flavor basis $(i,j = \nu_{e}, \nu_\mu, \nu_\tau)$. The propagation is governed by the equation,
\be 
 \frac{d {\bm{\rho}}}{d r}=-i\commute{\bf{H}}{\bm{\rho}} - \left.\frac{d\bm{\rho}}{d r}\right|_{NC,CC}
,
\label{eq:prop}
\ee
where $r$ is the distance from the center of the Sun. The $\frac{d\bm{\rho}}{d r}$ term denotes the neutrino flux attenuation from neutral-current (NC) and charged-current (CC) scatterings off the solar matter, and `re-injection' at lower energy after NC scattering, as well as secondary $\nu_e, \nu_{\mu}$ production from $\tau$ regeneration; see Appendix~\ref{app:prop} for details. As neutrinos are created in gauge eigenstates, $\bm{\rho}$ is a diagonal matrix at the center of the Sun with diagonal elements $\rho_{ii}$ denoting the fractions of the corresponding flavor. \ck{The flavor-basis Hamiltonian includes a rotation from a diagonal mass basis matrix and a matter effect term due to CC scattering with electrons~\cite{ref:msw}},
\be 
{\bf H}=\frac{1}{{2 E_\nu}}{\bf V}\,{\bf diag}\left(0, \delta m^2_{21},\delta m^2_{31}\right){\bf V^\dagger} \pm \sqrt{2} G_F n_{e} 
{\bf diag}(1,0,0)\,,
\ee
where $n_e$ is the electron number density, \ck{$G_F$ is the Fermi constant},
 and $\delta m^2_{ij}=m_i^2-m_j^2$ are the neutrino mass-squared differences. The sign of the matter term is positive for neutrinos and negative for antineutrinos. $\bf{V}$ is the neutrino mixing matrix
parametrized by three mixing angles $\theta_{ij}$ and a CP phase. We set the oscillation parameters to be~\cite{Maltoni:2004ei}:
\be 
\begin{array}{l}
\delta m^2_{21}= 8.1\times10^{-5}\, \text{eV}^2\,,\ \ \ \delta m^2_{31}= 
 2.2\times10^{-3}\,\text{eV}^2\,,\ \ \ 
\theta_{12}= 33.2\degree\,,\ \ \ \theta_{13}= 0\,,\ \ \  \theta_{23}=45\degree\,. \\
\end{array}
\ee

\begin{figure}
\includegraphics[scale=0.75]{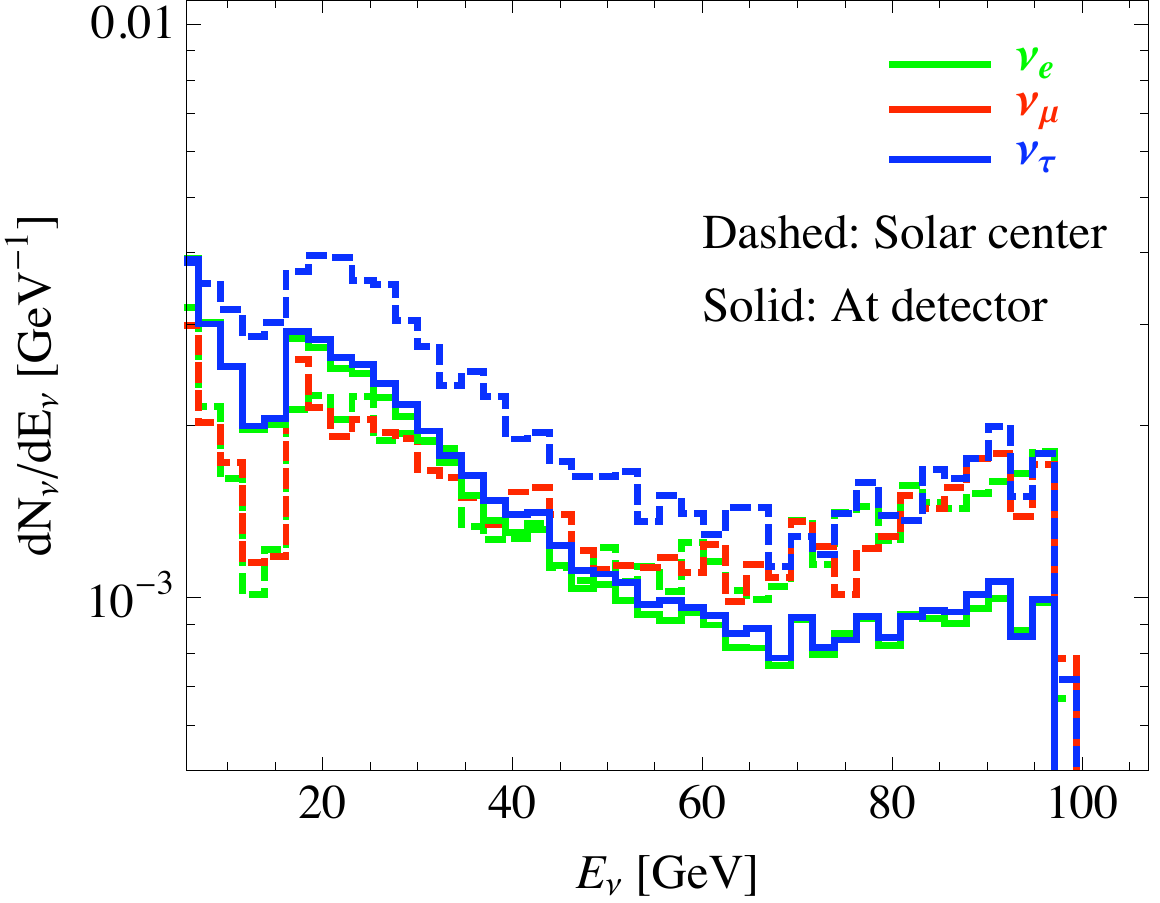}
\caption{Neutrino spectra for \ck{Point A} at the center of the Sun (dashed) and at the detector (solid). The dominantly transverse $WW$ channel leads to a hard spectrum. The attenuation is mainly due to CC scattering in the solar medium. NC scattering and $\tau$-regeneration feed neutrinos back to lower energy. The spectra at the detector are time-averaged by removing vacuum oscillatory components. The $\nu_\mu$ and 
$\nu_\tau$ spectra at the detector are almost identical. }
\label{fig:prop}
\end{figure}

After leaving the surface of the Sun, neutrino propagation is dictated by the vacuum Hamiltonian 
$\left.H\right|_{n_e=0}$. Our choice $\theta_{13}=0$ causes the very long wavelength modes to be suppressed so that IceCube measures the average neutrino flux in half a year. The vacuum-oscillation average is obtained by dropping the off-diagonal terms of $\bf{V}\bm{ \rho}$, {\it i.e.}, the neutrino density matrix in the mass basis. We ignore the attenuation of neutrinos as they pass through the Earth. Due to the low density of earth matter, attenuation becomes significant only above $10^5$ GeV and is negligible for the atmospheric background and sub-TeV DM neutrino sources. Fig.~\ref{fig:prop} shows propagation effects on the neutrino spectra for Point A.

\section{Detector Simulation}
\label{sec:detector}

The IceCube detector is a km$^3$ sensor array that tracks muons. By selecting events that come from below the horizon, the only source of muons are the atmospheric and cosmic neutrinos that penetrate the bulk of the Earth and CC scatter with nuclei inside or in the vicinity of the detector. The detected muons are grouped into two categories: `contained' muons with tracks starting within the instrumented volume, and `up-going' muons that are created under the detector and range into the detector~\cite{Erkoca:2009by}. DeepCore is an extension with six additional strings inside the IceCube array. DC has a muon detection threshold as low as 10~GeV and vetoes all the muons detected by the surrounding IceCube strings, thus eliminating most of the huge downward muon background and allowing 4$\pi$ detection of contained muons events. Recently Ref.~\cite{ref:cascades} pointed out that cascade events may be detectable and may increase the event count at IceCube significantly. However, the angular resolution of cascade events is much poorer than track-like muon events, leading to a significantly larger acceptance cone size for atmospheric neutrinos. We found that the atmospheric cascade event rate completely overwhelms the solar DM signal by a factor of 
$\sim10^2$ even with an optimistic angular cone size of $30\degree$. Consequently, we limit our study to track-like events. 

The muon threshold energy has a significant effect on the background rates. Although IC and DC can detect muons with energy as low as 50 and 10 GeV respectively, the angular reconstruction for the muon track requires triggering of least three optical modules which raises the energy threshold. In most of what follows, we assume the threshold energy to be 100~GeV and 35~GeV for IC and DC, respectively,
and  the angular resolution to be 1$\degree$ half apex angle. 

The contained muons are detected at their initial (maximum) energy and the rate is given by
\be 
\label{eq:contained}
\frac{d\phi_\mu}{dE_\mu d\Omega}=V(E_\mu)\eta(\theta_{z})\int_{E_\mu}^{E_\nu^{max}} dE_{\nu} \sum_{i=\nu_{\mu},\bar{\nu}_{\mu}}n_{n/p}\frac{d\sigma^{n/p}_i(E_\nu ,E_\mu)}{dE_{\mu}} \frac{d\phi^i_{\nu}}{dE_\nu}\,,
\ee
where $\frac{d\sigma^{n/p}}{dE_\mu}$ is the differential cross section of creating a muon of energy $E_\mu$ from CC scattering off a neutron/proton. $n_{n/p}$ is the numerical density of 
neutrons/protons in the medium and $\eta(\theta_{z})$ is the detection efficiency at zenith angle 
$\theta_{z}$. Note that at the South Pole the Sun stays within the range of $0\sim 23\degree$ from the horizon. We optimistically assume that the efficiency has a weak angular dependence and set 
$\eta=1$. This gives the maximum muon count which can be further adjusted with realistic detector information.  $\frac{d\phi_{\nu}}{dE_\nu}$ is the incoming neutrino flux at the detector.  $V$ is the detector volume which we take to be an energy-independent $1~\text{km}^3$ for  IC.
For DC we parametrize the effective volume at SMT3-trigger level (DC veto)~\cite{bib:deyoung} as
\be 
V_{\text{DC}} =\left\lbrace \begin{array}{cl}
0.32+7.54x+8.91x^2&, \ \ 0< x\leqslant 1.09  \\
11.9+7.77x-1.06x^2&, \ \ 1.09<x\leqslant 3.0 \\
26.1&,\ \  x > 3.0
\end{array}
\right.
\label{eq:DCvol}
\ee
where $x\equiv\loge$ and $V_{\text{DC}}$ is in megatons of water. The effective volume in ice is
 $V_{\text{DC}}^{\text{ice}}=V_{\text{DC}}\cdot \rho_{\text{water}}/\rho_{\text{ice}}$.

The up-going muons lose energy before reaching the detector modules. Muons with energy below a TeV lose energy mainly via ionization. Above a TeV, radiative losses becomes significant. Following Ref~\cite{GonzalezGarcia:2009jc}, the up-going muon rate is given by,
\bea 
\frac{d\phi_\mu}{dE_\mu d\Omega}&=&A_{\mu}(E_\mu,\theta_{z})\int_{0}^{\infty}dz\int_{E_\mu}^{E_\nu}dE^0_\mu \ P(E_\mu^0,E_\mu;z) \nn \\ 
&&\int_{E_\mu}^{E^{max}_{\nu}}dE_\nu \sum_{i=\nu_\mu, \bar{\nu}_\mu}n_{n/p}\frac{d\sigma^{n/p}_i(E_\nu ,E^0_\mu)}{dE_{\mu}^0} \frac{d\phi^i_{\nu}}{dE_\nu}\,,
\label{eq:upgoing}
\eea
where $E_\mu^0$ and $E_\mu$ denote the muon energy before and after propagating a distance $z$ outside the detector. 
$P(E_\mu^0,E_\mu;z)$ is the survival probability of the muon after propagation which we simulate using the Muon Monte Carlo package~\cite{Chirkin:2004hz}; for details see 
Appendix~\ref{app:mmc}. $A_{\mu}(E_\mu,\theta_{z})$ is the effective muon detection area~\cite{GonzalezGarcia:2009jc} parametrized by
\bea
\ \ \ \ \ \ A_{\mu}&=&1\text{km}^2\cdot A_0(E_\mu)\ (0.92-0.45\,\text{cos}\,\theta_{z}) 
\label{eq:affarea}\\
A_0&=&\left\lbrace
\begin{array}{cl}
0&,\ \ x < 1.6\\
0.784(x-1.6)&,\ \  1.6\leqslant x < 2.8\\
0.9+0.54(x-2.8)&,\ \  x \geqslant 2.8\\
\end{array} \right. \nn 
\eea where $x\equiv \text{log}_{10}(E_\mu\text{/GeV})$.

\begin{figure}
\includegraphics[scale=0.64]{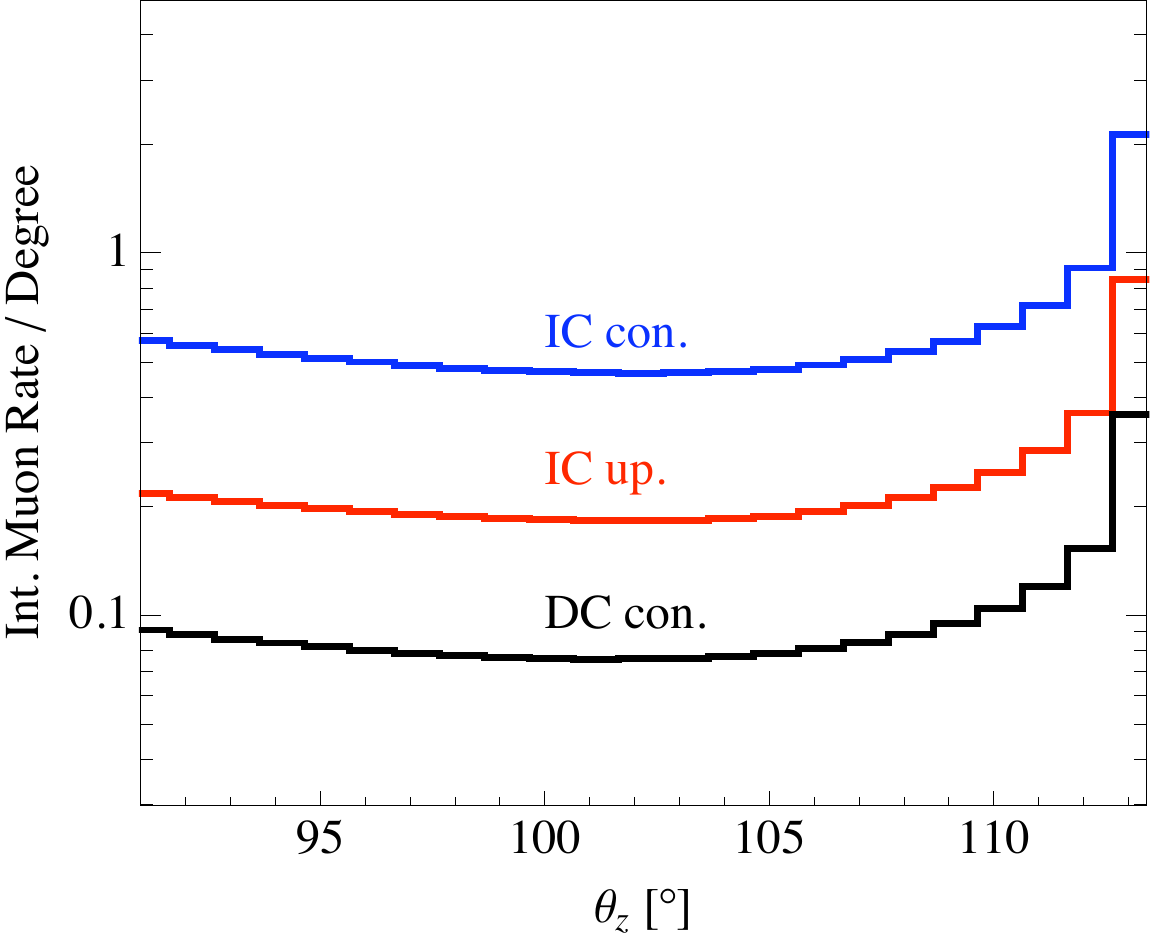}\hspace{0.5cm}
\includegraphics[scale=0.65]{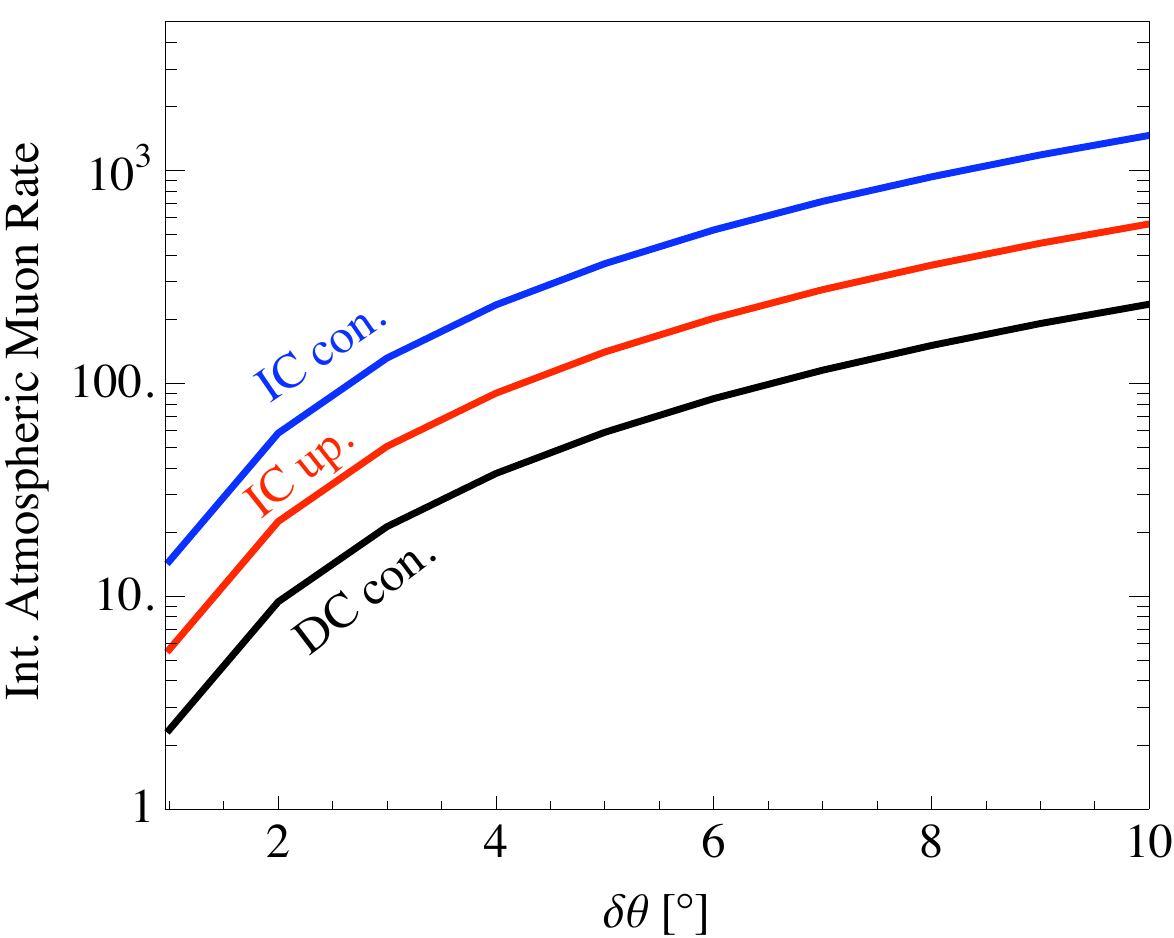}
\caption{Left panel: Integrated atmospheric background rate as the Sun sweeps through each degree in zenith angle in half-a-year at IceCube and a full-year at DeepCore assuming a 1$\degree$
acceptance cone. Right panel: The acceptance cone-size $\delta \theta$ (half the apex angle) dependence of the yearly atmospheric background rates.}
\label{fig:angbackgroud}
\end{figure}

Observation of up-going events at IC starts with the September equinox ($t=0$) and ends with the March equinox ($t=0.5$). During this interval, the zenith angle of the Sun follows
\be 
\theta_{z}(t)=90\degree+ 23.43\degree \text{sin}(2\pi t)  \ \ \ \ (0\leqslant t\leqslant 0.5 )\,.
\label{eq:height}
\ee
For IC, the muon rates are obtained by integrating Eq.~\ref{eq:contained} and Eq.~\ref{eq:upgoing} over the real-time zenith angle of the Sun; DC operates year-round so that observation time doubles. While the solar core is a point-like source, the major background for the DM signal comes from atmospheric neutrinos, which depends on the angular resolution of the detector. The directional atmospheric neutrino flux is measured by Super-K~\cite{Honda:2006qj} and is symmetric about the horizon $(\theta_{z}=90\degree)$. The left panel of 
Fig.~\ref{fig:angbackgroud} assumes a cone of 1$\degree$ (half the apex angle) and shows the muon rate from atmospheric neutrinos at IceCube and DeepCore as a function of zenith angle. From the right panel it should be noted that the background rate increases quadratically with the opening angle of the acceptance cone. For a \ck{more} detailed description of up-going and contained atmospheric background events, see Appendix C of Ref.~\cite{Barger:2010ng}.

\section{Dark matter signals}
\label{sec:signals}

\begin{figure}
\includegraphics[scale=0.8]{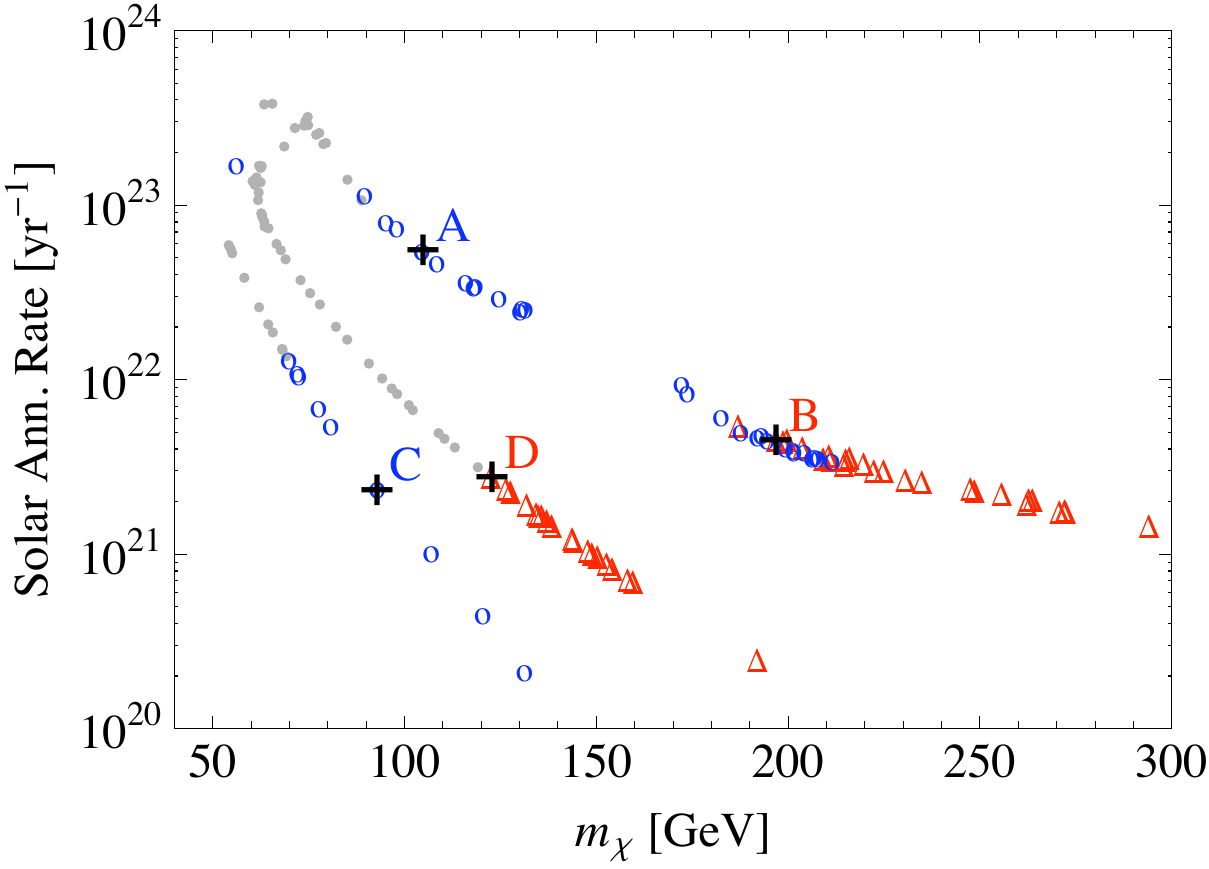}
\caption{\ck{Annihilation rate for mSUGRA points that are consistent with the measured relic density at the 95\% C.~L. Plotted points are from two scans in the ($m_{0}$,~$m_{1/2}$) plane with $\tan\beta$=10 and 50.  The light gray dots represent parameter points excluded by XENON100/Super-K/IC data.} \ck{Points A$-$D of Table~\ref{tab:susy} are marked among the nonexcluded points with $\tan\beta$=10 (blue circles) and $\tan\beta$=50 (red triangles).}
}
\label{fig:ann_rate}
\end{figure}

The number of energetic neutrinos above detector threshold determine the prospects for detecting new physics at IceCube/DeepCore. For a solar WIMP signal there are three major factors: (\textit{i}) annihilation rate; (\textit{ii}) muon energy threshold vs. WIMP mass; (\textit{iii}) annihilation channels that produce energetic neutrinos.

The relic density provides a stringent constraint that relates the first two factors. 
Fig.~\ref{fig:ann_rate} illustrates the dependence of the yearly annihilation rate on the neutralino mass in different regions of mSUGRA parameter space. The inverse dependence of the annihilation rate on WIMP mass is evident with a minimum mass set by the energy threshold of the
detector. Viable mSUGRA points must also be consistent with the XENON100~\cite{Aprile:2010um} constraint on $\sigma_{SI}$, \ck{which is the most stringent among nuclear recoil experiments~\cite{Szelc:2010zz}},  as well as the Super-K~\cite{Desai:2004pq} and IC~\cite{Abbasi:2009uz} constraints on $\sigma_{SD}$. 
The lower energy threshold of DeepCore greatly enhances the signal with respect to the background and can be crucial for WIMPs with mass $\sim$10$^2$ GeV.
The Focus Point region allows a large $\sigma_{SD}$ coupling and is the most popular discovery scenario for mSUGRA. 
The $b\bar{b}$ channel generally produces less and softer neutrinos compared to the $WW$ and $ZZ$ channels. The $\tau^+\tau^-$ channel provides energetic neutrinos but often has a low branching fraction. The neutrino signal from a massive WIMP becomes hard to detect as the $b\bar{b}$ channel dominates (Point D).

\begin{table}
\begin{tabular}{c|c|c|c|c|c}
\hline
Source &IC up.&IC con.&IC up.&IC con.&\hspace{0.2cm} DC\hspace{0.2cm}\\
\hline
 $E^{\mu}_{\text{thr}}$ (GeV)& 100 & 100 &70&70& 35\\
\hline
Atm. bkg.&5.6&14& 6.1  & 21 &2.3 \\ 	
A & \sci{1.8}{-4} & 0.042 & 8.2 & \sci{9.7}{2} &196 \\
B & 2.4& 66& 5.4 & \sci{1.7}{2} & 21\\
C &0 &0 & 0.016 & 2.9 &2.2 \\
D &0.011 &1.3& 0.18 & 14 & 3.2\\
\hline
\ck{E} & \sci{3}{-5} & \sci{4}{-3} & \sci{6}{-4} & 0.05 & 0.011 \\
F& 0 & 0 & 0 & 0 & 4.3 \\
G& 0 & 0 & 0 & 0 & $\sim 10^{-6}$\\

\hline
\end{tabular}
\caption{Atmospheric background rate and signal rates for the points of Table~\ref{tab:susy}. The observation time for IC and DC are $\frac{1}{2}$ and 1 year, respectively. The acceptance cone has a 1$\degree$ opening angle (half the apex angle) for both background and signal.  E$^{\mu}_{\text{thr}}$ denotes the muon energy threshold.}
\label{tab:intrates}
\end{table}

\begin{figure}
\includegraphics[scale=0.7]{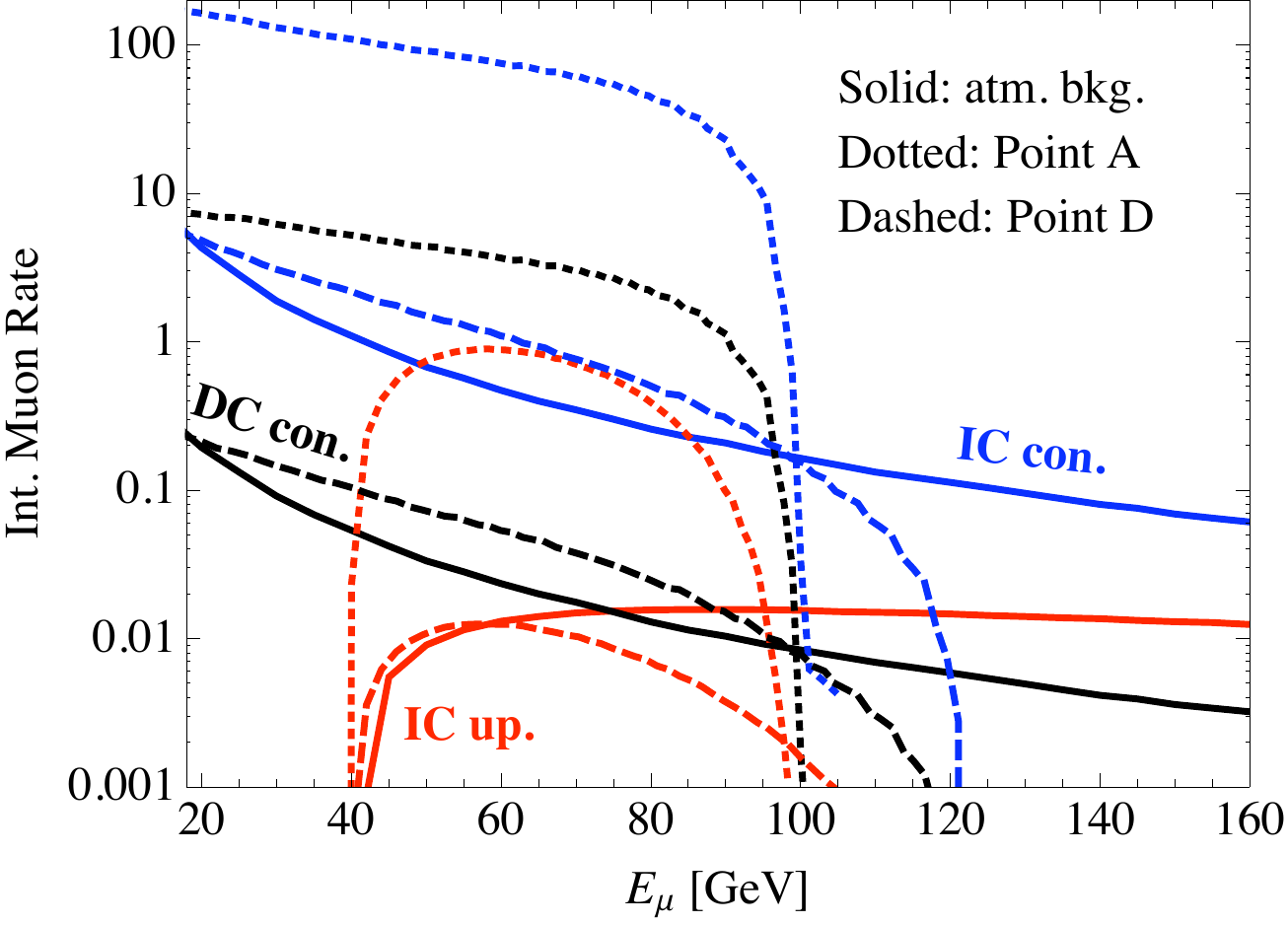}

\caption{Time-integrated muon energy spectra for the atmospheric background and WIMP signals. }
\label{fig:diffspec}
\end{figure}

In Table~\ref{tab:intrates}, we list the signal rate for the points in Table~\ref{tab:susy}, and the atmospheric background rates; for IC we have entertained two possible detector thresholds for comparison. 
The differential energy spectra for Points A and D are shown in Fig.~\ref{fig:diffspec}. With sufficient statistics as for Point A, it is even possible to construct the shape of the energy spectrum.

\section{Summary}
\label{sec:res}

We presented a calculation of neutrino source spectra from solar dark matter annihilation, accounting for spin-correlations, and a simulation of the propagation and detection of these neutrinos at the IceCube/DeepCore detector. We considered the mSUGRA model to illustrate spin-correlations, but similar techniques can be applied to other WIMP models.

With an angular resolution of $1\degree$ half apex angle and muon energy threshold of 100~GeV and 35~GeV for IceCube and DeepCore respectively, the yearly atmospheric background rate is 5.6 for IC upgoing events, 14 for IC contained events and 2.3 for  DC contained events. Generically, a neutralino annihilation rate of $~10^{21}$ yr$^{-1}$ is necessary to compete with the atmospheric background. With its lower energy threshold DC plays an important role in detecting neutrinos from relatively light ($\sim 10^2$ GeV) dark matter candidates, which have less suppressed annihilation rates. For DM masses that are not much above the detector
threshold, accounting for the helicity distribution of the final state particles can be critical for
the detectability of the signal. For example, $W$ pairs produced from neutralino annihilation are
transversely polarized and give
a much harder neutrino spectrum than if both helicities contribute equally. 

The neutrino-copious $W^+W^-, \tau^+\tau^-$ and $t\bar{t}$ channels are the main contributors to the neutrino signal. Since neutralinos are Majorana fermions, spin correlation requires $W^+W^-$ to be transversely polarized which yields a hard neutrino spectrum and enhances the muon rate above IC/DC thresholds. 
The Focus Point region is the primary mSUGRA discovery region for IC/DC because of the high annihilation rate. In general, the neutrino signal rate can be compromised if the annihilation occurs primarily into channels that do not produce a significant neutrino flux or a hard spectrum. 

\section*{Acknowledgments }

We thank Dmitry Chirkin, Tyce DeYoung, Arif Erkoca, Darren Grant, Francis Halzen, Jason Koskinen, Ina Sarcevic and Gabe Shaughnessy for useful discussions, and especially Enrico Maria Sessolo for many discussions and inputs.  This work was supported by the DOE under Grant Nos. DE-FG02-04ER41308, DEFG02-95ER40896 and DE-FG02-96ER40969, by the NSF under Grant No. PHY-0544278, and by the Wisconsin Alumni Research Foundation.
\bigskip

\appendix
\label{appendix}

\section{Source neutrino spectrum}
\label{app:spin}

The primary channels that contribute neutrinos are given in Table~\ref{tab:ch}. Pre-shower spectra are generated with MadGraph/MadEvent (MG/ME)~\cite{Alwall:2007st} or Calchep~\cite{bib:calchep} that keep the spin-correlation of the final state particles.
 We modified the MG/ME-Pythia~\cite{ref:pythia-pgs, Sjostrand:2006za} interface to develop the shower and take into account the solar absorption of $b,c$ hadrons. Secondary neutrinos come from subsequent decays of taus and $b, c$ hadrons in the shower.  In the case of $\tau$ decay, helicity information is facilitated by the Tauola~\cite{ref:tauola} package, as part of the MG/ME-Pythia interface.

In the $W^+W^-$ and $t\bar{t}$ channels, the undecayed $W$ and $t$ are made stable in the shower and their contribution is added as the charge conjugate of the neutrino spectra. Similarly, in the $ZZ$ channel one $Z$ boson is tagged stable in the shower and the neutrino spectrum is doubled. The $\tau^+\tau^-$ channel needs special treatment as a limitation of the LHE format~\cite{Alwall:2006yp} makes it difficult to pass the $s/t/u$ channel particle information to Pythia, and Tauola cannot reconstruct the helicity of the pair-produced $\tau$ leptons. We circumvented this problem by treating $\tau$ decay analytically for the $\tau^+\tau^-$ channel, as discussed in Appendix~\ref{app:tau_decay}. 

\begin{table}
\begin{tabular}{c|c}
\hline
\ \ Channel\ \ 	&	\ \ Final state	\ \ \\
\hline
$W^+W^-$	&	$W^{+*}\bar{\nu}_\tau \tau^-$,   $W^{+*}s\bar{c}$,   $W^{+*}d\bar{c}$	\\
$t\bar{t}$	&	$t^*\bar{b}\bar{\nu}_\tau \tau^-$,   $t^*\bar{b}s\bar{c}$,   $t^*\bar{b}d\bar{c}$	\\
\hline
$Z Z$	&	$Z^*\tau^+\tau^-$,   $Z^*\nu_l^+\nu_l^-$,   $Z^*b\bar{b}$, $Z^*c\bar{c}$	\\
$Z h$	&	$h^*\tau^+\tau^-$,   $h^*\nu_l^+\nu_l^-$,   $h^*b\bar{b}$, $h^*c\bar{c}$,	\\
  	    &	$Z^*\tau^+\tau^-$,   $Z^*b\bar{b}$,  $Z^*c\bar{c}$	\\
\hline
$\tau^+ \tau^-$ &	$\tau^+\tau^-$ \\
$b \bar{b}$	&	$b\bar{b}$	\\
$c \bar{c}$  &	$c\bar{c}$	\\
\hline
\end{tabular}
\caption{Neutralino annihilation channels that contribute neutrinos. Particles marled with a $*$ are made stable in the shower. The spin correlation in the $\tau^+\tau^-$ channel is treated separately with helicity-dependent decays.}
\label{tab:ch}
\end{table}
 
MadGraph has difficulty in producing transverse-$W$ spectra for a few mSUGRA points in the static limit and we switched to Calchep for the $W^+W^-$ channel. A drawback is that Calchep sums over final spins, so helicity information of the $\tau$ leptons is lost. Thus, most of the secondary (soft) $\nu_\tau$ component in the $W^+W^-$ channel is obtained from unpolarized $\tau$ decays. In the $t\bar{t}$ channel, final state radiation is turned off in Pythia to avoid a problem of flavor sum of parton clusters, after $t$ is tagged stable. The resultant $b$ quark energy may be high by a few percent. However, the effect is irrelevant since secondary neutrinos are dominantly produced from
$\tau$ decay.

\section{$\tau$ decay}
\label{app:tau_decay}

\begin{figure}
\includegraphics[scale=0.7]{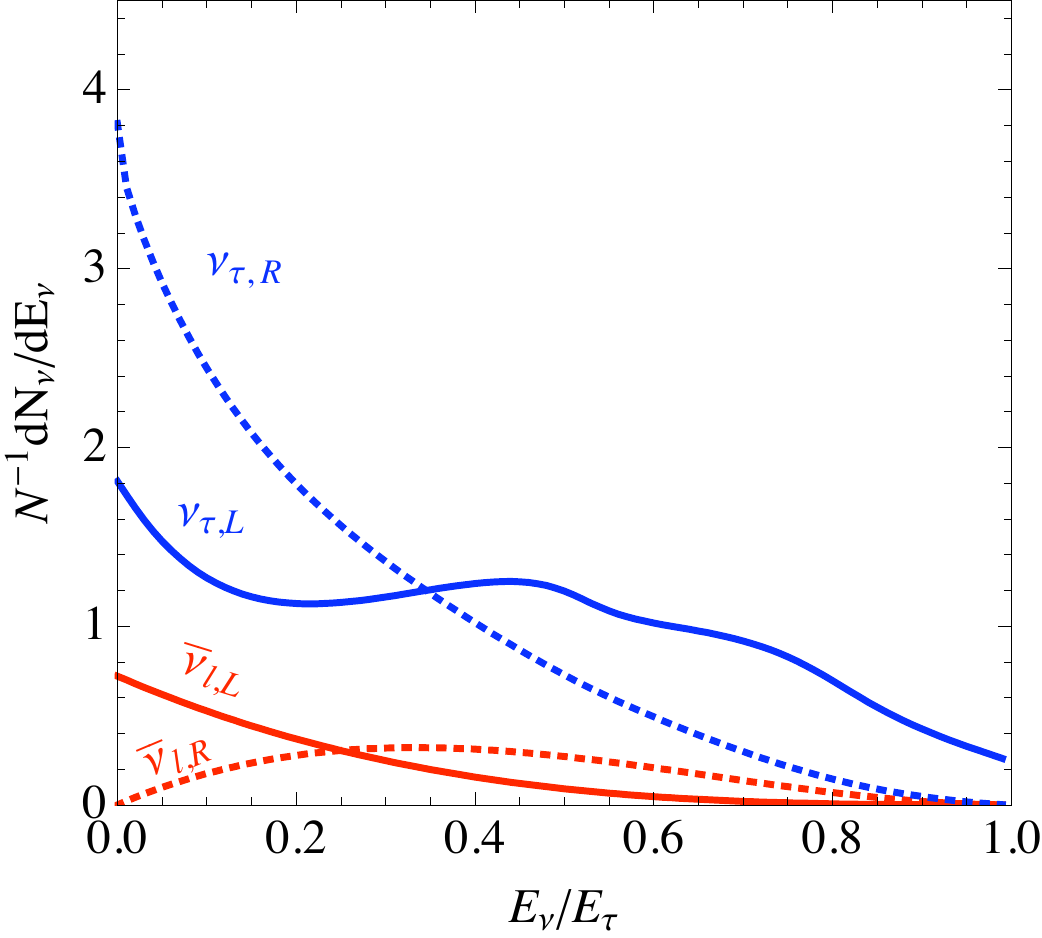}
\caption{$\nu_\tau, \bar{\nu}_l$ distribution from the decay of left(L)/right(R) handed taus.}
\label{fig:tauspec}
\end{figure}

While subdominant when the $W^+W^-$, $ZZ$ and $t\bar{t}$ channels are kinematically allowed, the $\tau^+\tau^-$ channel is a major source of neutrinos for lighter DM as the other dominant channel, $b\bar{b}$, produces less neutrinos per annihilation. The relevant decay channels are $\tau \rightarrow \nu_\tau l \bar{\nu}_l$ and $\nu_\tau$+ hadrons as listed in Table~\ref{tab:taudecay}.
The neutrino energy spectrum in the rest frame of the $\tau$ can be expanded as
\be
\frac{1}{N_0}\frac{dN}{dx d\text{cos}\theta}=f_0(x)+ f_1(x)\cos\theta\,,
\ee
where $x=2E_\nu/m_\tau$ is the energy fraction, and $f_0,f_1$ are projections of the distribution on the first two spherical harmonics. After boosting into the lab frame the first two harmonic coefficients are
\bea 
g_0(y)&=&\int_{y}^1 dx f_0(x)/x\,,\nn \\
g_1(y)&=&\int_{y}^1 dx (2y-x)f_1(x)/x^2\,,
\eea
where $y=E_\nu/E_\tau$. The lab frame neutrino spectrum is
\be 
\frac{1}{N_0}\frac{dN}{dy}=g_0(y)+P g_1(y)\,,
\ee 
where $P=\pm 1$ for a left/right-handed $\tau$.  For a two-body decay $f_{0,1}=\delta(1-x-m^2_X/m_{\tau}^2)$, while for the $\tau \rightarrow \nu_\tau l \bar{\nu}_l$ channel, these functions 
are~\cite{Lipari:1993hd}
\be
f_0 =\left\lbrace
	\begin{array}{cl} 
	2x^2(3-2x)&\ \ \text{for }\nu_\tau\\
	12x^2(1-x)&\ \ \text{for }\bar{\nu}_l\\
	\end{array}\right.
\,,\hspace{1cm}
f_1 =\left\lbrace
	\begin{array}{cl} 
	-2x^2(2x-1)&\ \ \text{for }\nu_\tau\\
	12x^2(1-x)&\ \ \text{for }\bar{\nu}_l\\
	\end{array}\right.\,, \nn
\ee
 and the lab-frame spectra are listed in Table~\ref{tab:taudecay}. For the short-lived mesons $\rho$ and $a_1$ we smear the Dirac-$\delta$ to account for their widths. The rest-frame distributions are modified with a Breit-Wigner approximation,
\be 
f_{0,1}=\delta(1- x-r_{mes})\rightarrow f_{0,1}^*=C\frac{1}{(1- x-r_{mes})^2 + m_{mes}^2\Gamma_{mes}^2 m^{-4}_{\tau}}\,,
\label{eq:smeareddis}
\ee
where $r_{mes}\equiv m^2_{\rho, a1}/m^2_\tau$, $\Gamma_{mes}$ is the meson decay width and $C$ is a normalization factor. 
The neutrino spectra resulting from left and right-handed $\tau$ decays are shown in Fig.~\ref{fig:tauspec}.
 
\begin{table}
\begin{tabular}{|c|c|c|c|}
\hline
 Decay mode ($\nu_\tau$) &  BF  &$g_0(y)$ & $g_1(y)$\\
\hline
$\nu_\tau \ell \bar \nu_\ell$  & 0.18 &${5\over3}-3y^2+{4\over3}y^3 $&$ {1\over3}+{8\over3}y^3-3y^2$\\
$\nu_\tau \pi$ & 0.12 & ${1\over 1-r_\pi}\theta(1-r_\pi-y)$&$ -{2y-1+r_\pi \over (1-r_\pi)^2}\theta(1-r_\pi-y)$\\
$\nu_\tau a_1$ & 0.13 & $\int_y^1 f_0^*(x)x^{-1}dx$ & $\int_y^1 (y-2x)f_1^*(x)x^{-2}dx$\\
$\nu_\tau \rho$ & 0.26 & $\int_y^1 f_0^*(x)x^{-1}dx$& $\int_y^1 (y-2x)f_1^*(x)x^{-2}dx$\\
$\nu_\tau X$ & 0.13 & $\int_y^1 f_X(x)x^{-1}dx$& 0 \\
\hline
Decay mode ($\bar{\nu}_l$) & BF &$g_0(y)$&$g_1(y)$\\
\hline
$\nu_\tau \ell \bar \nu_\ell$ & 0.18 &$2-6y^2+4y^3$ &$-2+12y-18y^2+8y^3$\\
\hline
\end{tabular}
\caption{$\tau$ decay modes, their branching fractions and fragmentation functions $g_0$ and $g_1$. The smeared distributions $f^*$ are as in Eq.~\ref{eq:smeareddis}. We approximated $f_X(y)$ for the inclusive $\tau\rightarrow\nu_\tau X$ channel with a generic phase-space decay into 4 pions and $\nu_\tau$.}
\label{tab:taudecay}
\end{table}


\section{Hadron absorption}
\label{app:absorption}

Hadrons that contain $c,b$ quarks contribute to the neutrino flux through their weak decay modes. However, the dense environment at the center of the Sun shortens the mean free path of these hadrons and absorption effects become significant  when the mean free path is comparable to the decay length. We take the nucleon scattering cross sections of $c,b$ hadrons~\cite{Edsjo:1997hp} to be
\be 
\sigma(E)=\left\lbrace
\begin{array}{cl}
1.4\times10^{-30}\, \text{m}^2 &\ \ \text{for mesons} \\
2.4\times10^{-30}\, \text{m}^2 &\ \ \text{for baryons} \\
\end{array}\right.
\ee
and the mean free path is
\be 
\lambda_s(E)=\frac{1}{n_{\odot} \sigma(E)}\,,
\ee
where $n_{\odot}$ is nucleon number density at the center of the Sun.
The decay length is
\be 
\lambda_d(E)=c\tau \gamma(E)\,,
\ee
where $\gamma$ is the Lorentz boost factor, $c$ is light speed and $\tau$ is rest-frame life time.

\begin{figure}
\includegraphics[scale=0.73]{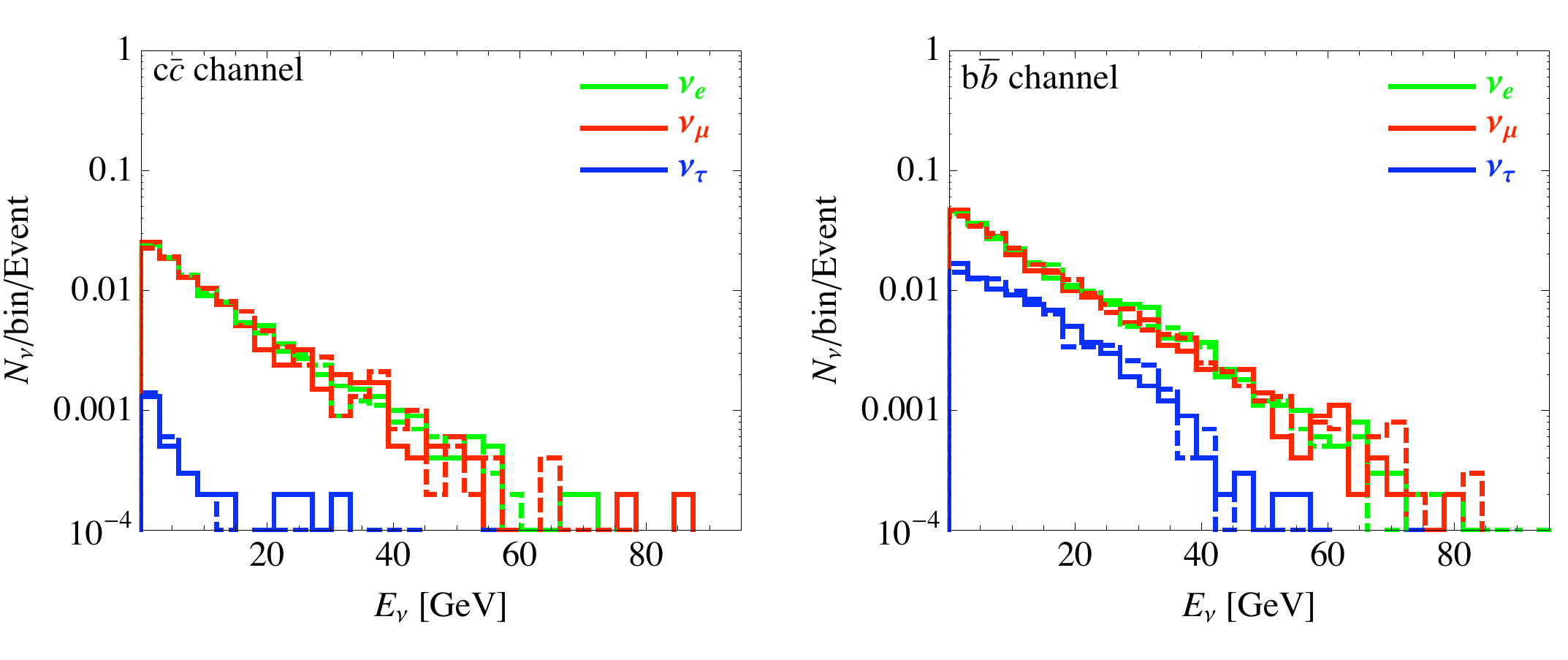}
\caption{Neutrino (solid) and antineutrino (dashed) spectra from the $c\bar{c}$ (left panel) and $b\bar{b}$ (right panel) channels for a 150 GeV WIMP. }
\label{fig:absorption}
\end{figure}

We assume that each scatter with a nucleus leaves the hadron with an average fraction  $\epsilon=0.7$ for $b$ hadrons, and $\epsilon=0.65\frac{m_c}{m_{had}}$ for $c$ hadrons~\cite{Ritz:1987mh} of the hadron's initial kinetic energy $E_0$. After the $n^{th}$ scattering, the ratio of the scattering probability to the decay probability is
\be
\frac{ P_{scattering}}{P_{decay}}=\frac{\lambda_d(E_n)}{\lambda_s(E_n)}\,,
\ee
where $E_n\sim E_0\epsilon^n$.
We modified the Pythia package to simulate the decay and energy loss of $c,b$ hadrons. The probability of scattering and decay are calculated iteratively for each hadron on an event by event basis. When a hadron is considered decayed, it is handed back to Pythia which develops the decay and showering. Since not all scatterings are elastic, we
allowed the hadrons to scatter only once before decaying or being stopped/destroyed. The resultant neutrino spectra for a 150 GeV WIMP are shown in Fig.~\ref{fig:absorption}.

\section{Neutrino scattering in propagation}
\label{app:prop}

The neutrino flux becomes attenuated by scattering off solar nuclei as it traverses the Sun. NC scattering decreases the neutrino energy while CC scattering converts the neutrino into the corresponding lepton whose decay introduces a secondary influx of less energetic neutrinos. The latter is often called ``regeneration". While electrons and muons are quickly absorbed by the dense solar medium, $\tau$ leptons decay promptly and contribute to the soft component of the neutrino flux. 

The flux attenuation caused by NC scattering is flavor blind,
\be 
\left.\frac{d \bm{\rho}(E_{\nu})}{dr}\right|_{NC}^{att.}=-\ n_{p/n} \sigma_{p/n}^{NC}(E_{\nu}) \bm{\rho}(E_{\nu}),
\ee
where $\bm{\rho}$ is the flavor density matrix and $n_{p/n}$ is the solar proton/neutron number density~\cite{Bahcall:2004yr}. The injection of scattered neutrinos is given by
\be 
\left.\frac{d \bm{\rho}(E_{\nu})}{dr}\right|_{NC}=n_{p/n} \int_{E_\nu}^{E^{max}_{\nu}} dE'_{\nu} \frac{d\sigma_{p/n}^{NC}(E'_{\nu},E_\nu)}{dE_{\nu}} \bm{\rho}(E'_{\nu})\,,
\ee
where the dummy $E_{\nu}'$ integrates over energy above $E_{\nu}$.

The attenuation via CC scattering is given by~\cite{Strumia:2006db}, 
\bea
\left.\frac{d \bm{\rho}(E_\nu)}{dr}\right|_{CC}^{att.}&=& 
- \frac{\anticommute{\bm{\Gamma}_{CC}}{\bm{\rho}}}{2}\,, \\
\text{where } \bm{\Gamma}_{CC}&=&{\bf diag}\left(n_{p/n}\ \sigma^{CC}_{p/n}(E_\nu)\right)\,, \nn
\eea
and the secondary contribution from $\tau$ regeneration is 

\bea 
\left.\frac{d \bm{\rho_{ij}}(E_{\nu})}{dr}\right|_{CC}^{reg.}&=&n_{p/n} \int_{E_{\nu}}^{E_{\nu}^{max}}dE_{\nu}'\int_{E_{\nu}}^{E_{\nu}'} 
\left.\frac{dE_\tau}{E_{\nu}}\right[  \\ \nn
&&\Pi_{\tau}\frac{d\sigma_{p/n}^{CC, \nu_\tau}(E_{\nu}',E_\tau)}{dE_{\tau}}g_{\nu_\tau}(E_{\nu}/E_\tau) \bm{\rho}_{\tau\tau}(E_{\nu}') \\ \nn
&+&
\left.
\Pi_{e,\mu}\frac{d\sigma_{p/n}^{CC, \nu_e,\nu_\mu}(E_{\nu}',E_\tau)}{dE_{\tau}}g_{\nu_{e,\mu}}(E_{\nu}/E_\tau) \bar{\bm \rho}_{\tau\tau}(E_{\nu}')
\right],
\eea  
\ck{where $\Pi_l=\delta_{il}\delta_{jl}$ is a $3\times3$ matrix with only the $l^{th}$ diagonal element nonzero and $l= 1,2,3$ for $e,\mu,\tau$ respectively. $\Pi_l$ picks out the diagonal terms from ${\bm \rho}$. $\bar{\bm \rho}$ denotes the density matrix for antineutrinos.} Note that since neutrinos have definite helicity, the $\tau$ leptons from CC scattering are polarized; $\tau^-$ is left-handed while $\tau^+$ is right-handed. See Appendix~\ref{app:tau_decay} for the helicity fragmentation functions $g_{\nu_l}(y)$ for $\tau$ decay. The neutrino-nucleon scattering cross sections are evaluated with the numerical package Nusigma~\cite{nusigma}.

\section{Muon propagation in ice}
\label{app:mmc}

The muon energy spectrum $\phi_{\mu}(E_{\mu},z)$ softens after propagation in ice, as generically described by
\bea
v_{\mu}\ \partial_z\phi_{\mu}(E_{\mu},z)&=&-\int_0^{E_{\mu}} dE_{\mu}'\phi_{\mu}(E_{\mu},z) n(z) \frac{d\sigma(E_{\mu},E_{\mu}')}{dE_{\mu}'} \nonumber \\ 
&&+\int_{E_{\mu}}^{E_{max}} dE_{\mu}''\phi_{\mu}(E_{\mu}'',z) n(z) \frac{d\sigma(E_{\mu}'',E_{\mu})}{dE_{\mu}''} +\partial_{E_{\mu}}(\alpha(E_{\mu})\phi_{\mu}(E_{\mu},t))\,,
\eea
where $v_{\mu}\approx c$ is the muon speed, $\sigma$ is the muon scattering cross section, 
$n(z)$ is the target density, and $\alpha$ describes ionization energy loses. $E_{\mu}$ is the muon's energy at a propagated distance $z$ and dummy variables $E'_{\mu}, E''_{\mu}$ denote the energy below/above $E_{\mu}$. In terms of the survival probability $P(E_{\mu}^0,E_{\mu};z)$, the final energy spectrum can be written as 
\be 
\phi_{\mu}(E_{\mu},z)=\int P(E_{\mu}',E_{\mu};z-z_0) \phi_{\mu}(E_{\mu}',z_0)dE_{\mu}'\,,
\ee
where $ E_{\mu}^0$ is the initial muon energy. \ck{For $z\neq 0$, $P(E_{\mu}^0,E_{\mu};z)$ can be below unity upon integration over $E_{\mu}$ due to muons being stopped before reaching $z$; $P(E_{\mu}^0,E_{\mu};z)$ can be obtained from Monte Carlo simulations.}

The spatial integral in Eq.~\ref{eq:upgoing} can be done separately to yield the so-called ``effective muon range"~\cite{GonzalezGarcia:2009jc},
\be
R(E_{\mu}^0,E_{\mu})=\int_0^{\infty}P(E_{\mu}^0,E_{\mu};z) dz\,.
\ee
$R(E_{\mu}^0,E_{\mu})$ represents the average incremental distance a muon travels per unit energy loss (in GeV) at $E_{\mu}$.
 
The average neutrino effective area is~\cite{GonzalezGarcia:2009jc},
\be 
A^{eff}_{\nu}(E_{\nu},\theta_z)=\frac{1}{2}\sum_{i=\nu_{\mu},\bar{\nu}_{\mu}} 
\int dE_{\mu}dE_{\mu}^0 n_{n/p}\frac{d\sigma^{n/p}_i(E_\nu ,E^0_\mu)}{dE_{\mu}^0} R(E^0_\mu,E_\mu) A_{\mu}(E_{\mu},\theta_z)\,,
\ee
where $A_{\mu}(E_{\mu},\theta_z)$ is the muon effective area given by Eq.~\ref{eq:affarea}. The attenuation of sub-TeV neutrinos inside the Earth can be ignored.

\begin{figure}
\includegraphics[scale=0.68]{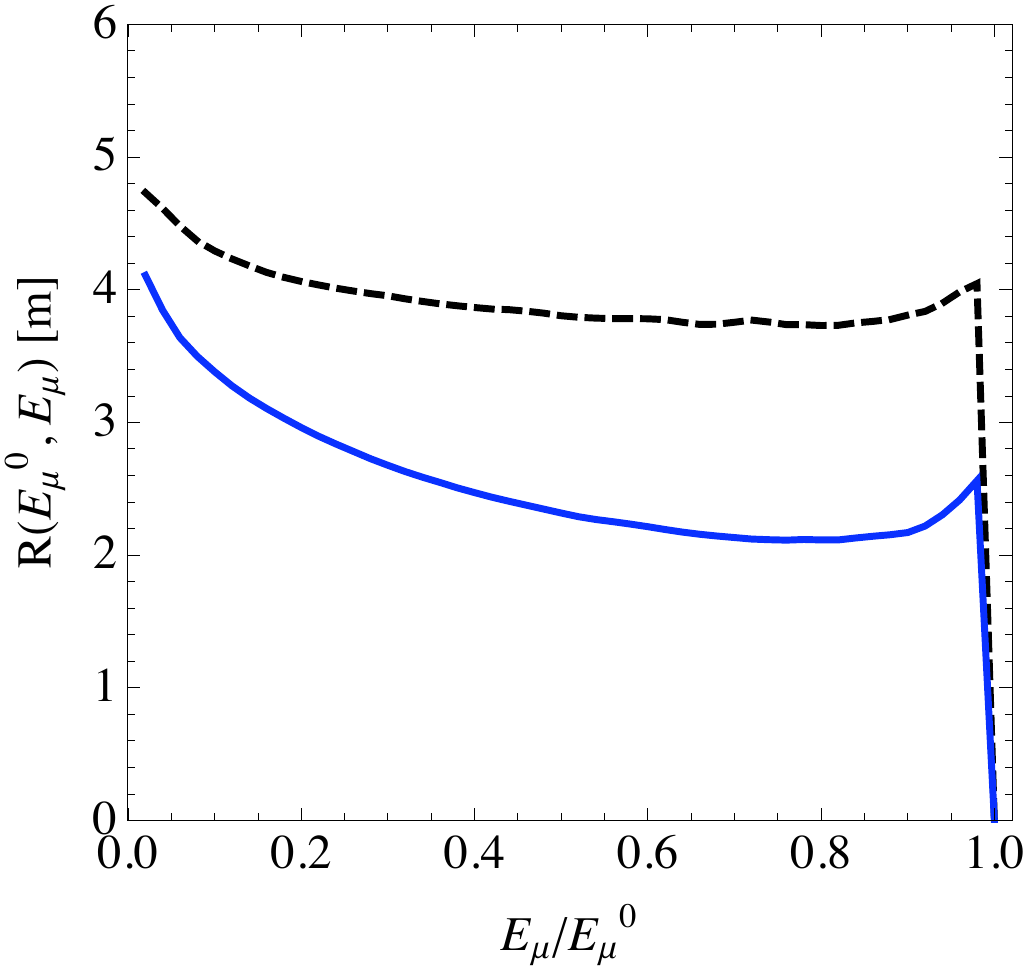}\hspace{0.5cm}
\includegraphics[scale=0.68]{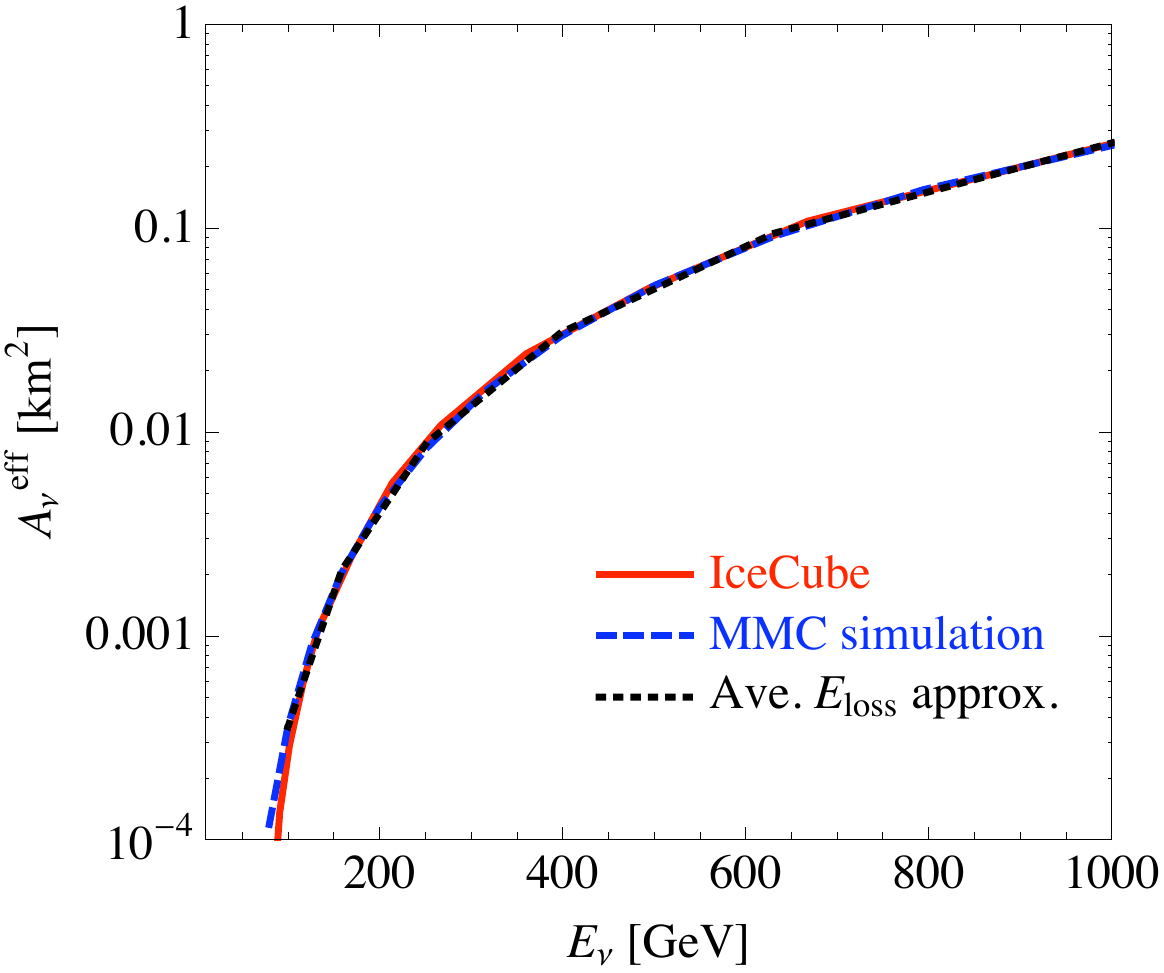}
\caption{Left pannel: 
\ck{Differential effective range $R(E_{\mu}^0,E_{\mu})$ in ice for  100 GeV (dashed) and 1 TeV (solid) muons.}
Right panel: Comparison of $A^{eff}_{\nu}$ from the IceCube detector 
simulation~\cite{GonzalezGarcia:2009jc}, our simulation with MMC, and the approximate method using a parameterization of average neutrino energy losses. The plotted $A^{eff}_{\nu}$ is an angular average over $\theta_z >90\degree$. With $\alpha=2.5\times 10^{-3}\text{ GeV\,cm}^{2}\text{/g}, \beta=4\times10^{-6}\text{ cm}^{2}\text{/g}$, the average $E_\nu$ loss method agrees well with sophisticated simulations.
}
\label{fig:effrange}
\end{figure}

We used Muon Monte Carlo (MMC) to simulate the muon survival probability $P(E_{\mu}^0,E_{\mu};z)$. The interpolated effective range for 100 GeV and 1 TeV muons are shown in Fig.~\ref{fig:effrange}. 
As a test we calculated  $A^{eff}_{\nu}$ and compared to the full detector simulation, and found excellent agreement as is evident from Fig.~\ref{fig:effrange}.

Alternatively, a frequently used approximation for muon propagation parameterizes the average muon energy loss,
\be 
\frac{dE_\mu}{dz}=-\rho(\alpha+\beta E_\mu)\,,
\ee
where $\rho$ is the medium density and $\alpha,\beta$ account for ionization and radiative effects. This procedure ignores the smearing of the muon energy distribution during propagation. For ice, we found $\alpha=2.5\times 10^{-3}\text{\ GeV cm}^{2}\text{/g}$, $\beta=4\times 10^{-6}\text{ cm}^{2}\text{/g}$ agrees well with the MMC results, as shown in Fig.~\ref{fig:effrange}.
There is a mild degeneracy in the parameters $\alpha$ and $\beta$, so that $\alpha=3\times 10^{-3}\text{\ GeV cm}^{2}\text{/g}$, $\beta=3\times 10^{-6}\text{ cm}^{2}\text{/g}$~\cite{Barger:2010ng}, works just as well.

\newpage

\end{document}